\documentclass[12pt]{article} 
\usepackage{cyr}
\usepackage{epsf}
\usepackage{pazh}

\usepackage{amsmath}
\usepackage{amsfonts}
\usepackage{amssymb}
\usepackage[english]{babel}
\usepackage{graphicx}
\graphicspath{ {./pictures/} }

\usepackage{multirow}
\usepackage{scalerel}
\usepackage[usenames]{color}
\usepackage{colortbl}
\usepackage{xcolor}
\usepackage[font={small}]{caption}

\newcommand\ddfrac[2]{\frac{\displaystyle #1}{\displaystyle #2}}

\tightenlines

\def\*{$^{*}$}
\def\Á{$^{\mbox{\small a}}$}
\def\Â{$^{\mbox{\small b}}$}
\def\Ũ{$^{\mbox{\small c}}$}
\def\Į{$^{\mbox{\small d}}$}
\def\Ä{$^{\mbox{\small e}}$}
\def\etal{{et~al.}}
\sloppypar

\begin{document}
\baselineskip 18pt

\noindent Astronomy Letters, v. 48, No. 12, pp. 689-701, 2022.

\title{\bf \Large The influence of the effective number of active and sterile neutrinos on the determination of the values of cosmological parameters.}

\author{\normalfont
P.A. Chernikov \affilmark{1*}, A.V. Ivanchik \affilmark{1*}}

\affil{
{\small \it \affilmark{1}Ioffe Instittute, Russian Academy of Sciences, Polytekhnicheskaya 26, St. Petersburg, 194021 Russia}}

\vspace{-2mm}
\noindent Accepted November 15, 2022

\sloppypar 
\vspace{-10mm}
\noindent
\title{\bf Abstract}
\vspace{-5mm}
Neutrino, as the second most abundant known particle in the Universe, has a significant impact on its expansion rate during the radiation- and matter-dominated eras. For this reason, a change in the number of neutrino species can lead to substantial changes in the estimates of cosmological parameters, the most accurate values of which, at the moment, are obtained by analysing the anisotropy of the CMB. In the presented work we consider the influence of a hypothetical sterile neutrino (with eV-scale mass) on the determination of cosmological parameters. The possible existence of such neutrino is suggested by the analysis of a series of different experiments. If it is detected, it will be necessary to include it into the $\Lambda \rm CDM$ model with the fixed values of its mass $m_{\rm s}$ and mixing angle $\theta_{s}$, which is the main method used through this paper. Apart from that, the seesaw mechanism requires there to be at least two sterile states, one of them being much heavier than the active ones. The heavier sterile state ($m_{s}\sim1$ keV) would decay and increase the effective number of active neutrinos. Therefore, the influence of a change in the effective number of relativistic neutrino species $N_{\rm eff}$ was studied as well, which could be caused by various reasons, for example, by the decay processes of dark matter particles or the above-mentioned sterile neutrinos, as well as processes leading to an increase in the temperature of relic neutrinos $T_{\scaleto{\rm C\nu B}{4pt}}$. The effects studied in this work lead to a significant change in the estimates of the cosmological parameters, including the value of $H_{0}$. It has been discovered that the accounting of the sterile neutrino with masses $m=1$ and $2.7$ eV leads to a decrease in the estimate of the current Hubble parameter value  $H_{0}$ and, therefore, exacerbates the ``$H_{0}$-tension'' problem. An increase in the value of the effective number of relativistic neutrino species leads, on the contrary, to an increase in the $H_{0}$ estimate, resolving the above-mentioned problem at $N_{\rm eff}=3.0+0.9$, which is equivalent to an increase of the neutrino temperature up to $T ^{\,0}_{\scaleto{\rm C\nu B}{4pt}}=1.95+0.14\,\rm K$. At the same time, the rest of the cosmological parameters do not change significantly, leaving us within the framework of the standard $\Lambda \rm CDM$ model.

\noindent
{\bf Keywords:\/} Cosmology, cosmic microwave background, sterile neutrinos, cosmological parameters, Hubble parameter, $H_{0}$-tension.

\vfill
\noindent\rule{8cm}{1pt}\\
{$^*$ e-mail: $<$filippcernikov@gmail.com$>$, $<$iav@astro.ioffe.ru$>$}

\clearpage

\section*{\large Introduction}
	\vspace{-5mm}
 Neutrino is the second most abundant known particle in the Universe, after photon.  Relic neutrinos\footnote{This refers to equilibrium relic neutrinos described by the Fermi-Dirac distribution. There also exist nonequilibrium relic neutrinos of primordial nucleosynthesis (see, for example, Ivanchik and Yurchenko 2018; Yurchenko and Ivanchik 2021).}, like CMB photons (or relic photons), are also formed during the first moments after the Big Bang. Within the framework of the Standard Model of particle physics, all three generations of neutrinos $(\nu_e, \nu_{\mu}, \nu_{\tau})$ are massless (Zyla, 2020). However, as a result of a series of experiments (collaboration ``Super-Kamiokande'', 1999 and ``SNO'', 2013), the effect of oscillations of the three  neutrino states among themselves was discovered, the cause of which, according to the theory developed by Bilenkiy and Pontecorvo (1976), is the presence of non-zero mass values for neutrinos. The established fact of oscillations thus confirmed the existence of a neutrino mass, but the nature of its occurrence, as well as the specific values of the masses, are still unknown.
 
There are various models for generating neutrino masses, some of them suggest the presence of additional neutrino states that do not participate in the interactions of the Standard Model, which are usually called sterile neutrinos, while the three other states are known as active. 

In the course of several experiments on the observation of reactor antineutrinos (Mueller et al. 2011, Mention et al. 2011, Gariazzo et al. 2017), a deviation of the predicted value of their flux from the observed one was revealed at different distances from the reactor. As one of the possible explanations for the effect, which has been called the ``antineutrino anomaly'', the process of antineutrino oscillations to a sterile state has been proposed.
Later, a number of other independent experiments in which the same effect was noticed, namely the collaboration ``NEOS'' (2017), ``DANSS'' (Danilov, 2020), ``STEREO'' (Allemandou et al., 2018) , ``PROSPECT'' (Ashenfelter et al., 2019) and ``BEST'' (Barinov et al., 2022) gave restrictions on the allowed range of sterile neutrino parameters ($\Delta m_{1 4}^2 \sim1~\text{eV}^2$, where $\Delta m_{1 4}^2$ is the difference between the squares of the lightest active state mass $m_{1}$ and $m_{4}$ of the sterile state). The ``Neutrino-4'' experiment (Serebrov et al., 2021), that specifically aimed at searching for a sterile state, gave the following estimate for the squared mass difference:
$\Delta m_{1 4}^2=7.3\pm1.17~\text{eV}^2$.
	
Expansion of the standard cosmological $ \Lambda \rm CDM $ model with the inclusion of a sterile neutrino will change the Universe expansion rate during two stages of its evolution: first in the radiation-dominated era, and then during the stage of nonrelativistic matter domination. This will affect both the processes of primordial nucleosynthesis and the formation of the CMB anisotropy pattern, and hence affect the most accurate (to date) values of the cosmological parameters obtained from them. The study of the effects of sterile neutrinos on the determination of cosmological parameters is one of the aims of this paper. Moreover, a change in the parameters of active states, namely in the effective number of neutrino species $N_{\rm eff}$, can also affect CMB anisotropy and the process of cosmological parameters determination. This is another purpose of the presented work.

\section*{\large Restrictions of primordial nucleosynthesis on neutrino properties}
\vspace{-5mm}
The most stringent restrictions on the neutrino properties (the number of generations of active neutrinos and the possibility of of a light sterile neutrino existence $(m_s \sim 1 - 3\,$eV)) are obtained from the analysis of primordial nucleosynthesis -- the abundance of $^4$He is most sensitive to the expansion rate of the Universe, which in turn is determined by the relativistic degrees of freedom during that epoch. Therefore, the addition of another active or light sterile neutrino may lead to a change in the $^4$He abundance that is not consistent with observational data, the most accurate of which have been obtained in recent works (Kurichin et al. 2021a, Kurichin et al. 2021b). Taking into account the ``Li-problem'', the variance of estimates for deuterium (see, e.g., Balashev et al., 2016), and the helium data, a conservative estimate allows one to consider no more than one additional relativistic degree of freedom.

\section*{\large The method for determining cosmological parameters based on the analysis of CMB radiation anisotropy}
\vspace{-5mm}

An analysis of the relic radiation temperature anisotropy (Planck Collaboration, 2020) allows estimates to be made for six key cosmological parameters: $\Omega_{\rm b},~\Omega_{\rm \text{\tiny CDM}},~\theta_{*},~n_{\rm s},~A_{\rm s},~\tau~$-- the present values of the relative density of baryons and cold dark matter, the angular acoustic scale, the scalar spectral index, the primordial comoving curvature power spectrum
amplitude, the optical depth of the reionized plasma. Using the obtained values of these six parameters, a number of other cosmological quantities can be determined. In our paper, in addition to the mentioned parameters, we will be interested in the present value of the Hubble parameter $H_{0}$ in the light of the recent discrepancy between its values obtained by two independent methods: 
one of them is the result of the CMB temperature anisotropy analysis (Planck Collaboration, 2020):
	\begin{equation}H_0=67.36\pm0.54 ~\text{km}\cdot\text{s}^{-1}\cdot \text{Mpc}^{-1} \end{equation}
	The other comes from observations of Type Ia supernovae (Riess et al., 2022):
	\begin{equation}
		H_0=73.04\pm1.04 \text{km}\cdot\text{s}^{-1}\cdot \text{Mpc}^{-1}
	\end{equation}
	These estimates differ by more than $5\sigma$, the discrepancy in these values has been called ``$\!H_{0}-$tension''. 
	
To analyze the anisotropy of the CMB, the distribution of its temperature deviations $\Delta T(\vec n)$ 
 from its current value $ T^{\,0}_{\rm \text{\tiny CMB}}=2.7255\pm0.0006\,K$ (Fixen, 2009) in the direction specified by the unit vector $\vec n$ on the celestial sphere is used:
	\begin{equation}\Delta T(\vec n)=T(\vec n)-T^{\,0}_{\rm \text{\tiny CMB}}\end{equation}
	The information about CMB temperature fluctuations is further presented in terms of coefficients $C_{l}$ obtained through decomposition of the correlation function for temperature values in two directions $\vec{n}$ and $\vec{n}'$ with respect to Legendre polynomials $P_{l}(\vec{n}\cdot\vec{n}')$ (for details, see, for example, the monograph of  Gorbunov et al., 2016, vol. 2): 
	\begin{equation}<\Delta T(\vec{n})\Delta T(\vec{n}')>=\sum_{l=0}^{\infty}C^{\,\scaleto{TT}{4pt}}_{ l}(\frac{2l+1}{4\pi})P_{ l}(\vec{n}\cdot\vec{n}')\end{equation}
	The values of the coefficients\footnote{The index $TT$ indicates that the correlation function of two temperature values is being considered. In addition to it, two more correlation functions associated with the polarization of CMB are often introduced. These are also used in the current work, but are not mentioned here for the sake of brevity.} $C^{\,\scaleto{TT}{4pt}}_{l}$ are, on the one hand, determined based on measurements of the CMB anisotropy (by inverting expression (4)), and, on the other hand, are determined from a numerical calculation of the evolution of $T^{\,0}_{\rm \text{\tiny CMB}}$ fluctuations. Comparison of the $C^{\,\scaleto{TT}{4pt}}_{l}$ values obtained by two different methods allows one to make constraints on the cosmological parameters used in the numerical simulation of CMB fluctuations. Figure 1 shows the dependences of $D_{l}^{\text{\tiny TT}}= \frac{l(l+1)}{4\pi}C_{l}^{\,\scaleto{TT}{4pt}}$ on $l$ obtained using this method under standard assumptions (Planck Collaboration, 2020) as well as including additional neutrino degrees of freedom. The values $D^{\,\scaleto{TT}{4pt}}_{l}$ retrieved from observational data are taken from the Planck Collaboration archive \footnote{This paper is based on observations obtained with Planck (http://www.esa.int/Planck), an ESA science mission with instruments and contributions directly funded by ESA Member States, NASA, and Canada. Planck collaboration archive link: https://pla.esac.esa.int}.
 
 To analyse the anisotropy of the relic radiation, the following assumption is often adopted, whereby the sum of neutrino masses has the lowest observationally admissible value (Planck Collaboration, 2020):
	\begin{equation}
	    m_{1}=m_{2}=0\,,~m_{3}\approx0.06~\text{eV}
	\end{equation}
	It is also used in this paper. In this case, the sterile neutrino in question has a mass of either $m_{4}\approx 2.7$ eV according to the ``Neutrino-4'' experiment, or $m_{4}\sim 1$ eV based on earlier estimates from reactor experiments.\\
 
 \pagebreak
	\begin{figure}[h!]
		\centering
		\includegraphics[width=\textwidth]{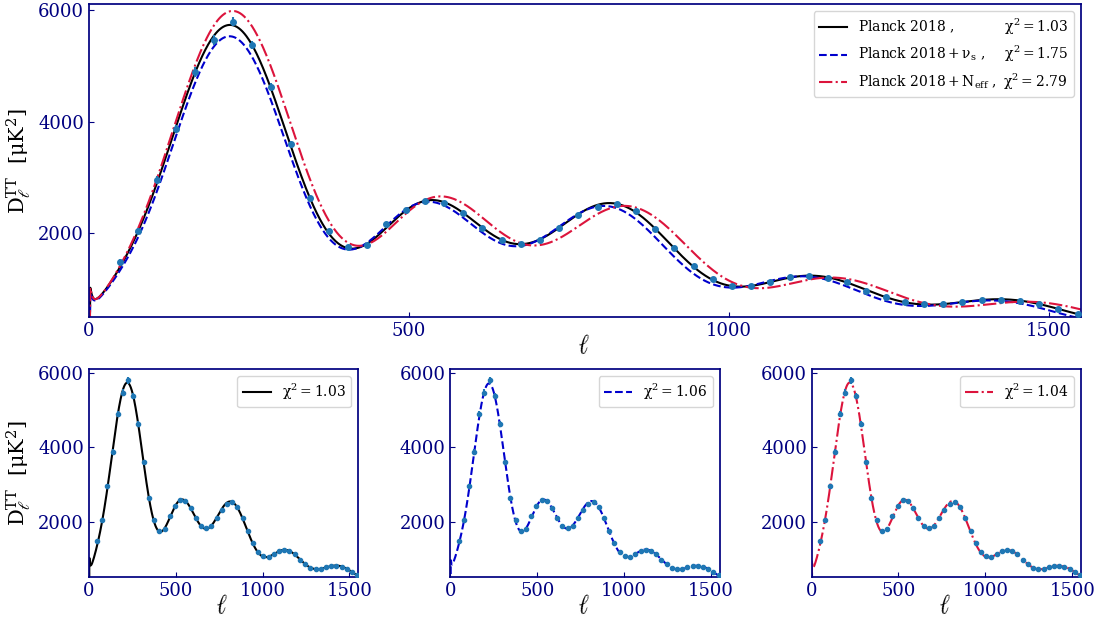}
		\caption{The multipole coefficients $D_{l}^{\text{\tiny TT}}$ that characterise the square amplitude of the CMB temperature fluctuations are represented in the figure by the blue dots obtained from the analysis of observational data. The lines (solid and dashed) in the upper panel correspond to the results of the theoretical calculations of $D_{l}^{\text{\tiny TT}}$: The black curve best approximates the observations within the standard $\Lambda \rm CDM$ model and determines the values of the main cosmological parameters. These values were also fixed when calculating the blue and red curves, which take into account the introduction of the sterile neutrino and additional degrees of freedom of active states, respectively ($N_{\rm eff}=3+1$ in both cases).
		It can be seen (among other things, from the value of the reduced $\chi^2$) that the introduction of additional number of neutrino species, both active and sterile, leads to a mismatch between the theoretical model and the observed data. However, as shown in the graphs of the bottom row, redefinition of the cosmological parameters can return the numerical model to fit the observations (first panel -- standard cosmological model, second panel -- addition of the sterile neutrino, third panel -- the increase in active neutrino degrees of freedom).}
	\end{figure}
	\pagebreak

\section*{\large Neutrino influence on the Hubble parameter determination}
\vspace{-5mm}

The expansion rate of the Universe is characterised by the Hubble parameter $\mathrm{H}(a)$, the dependence of which on the scale factor $a$ under the assumption of zero spatial curvature is given by the following expression
\footnote{The key cosmological equations used in this paper are presented, for example, in monographs by Weinberg (2008) and Gorbunov, Rubakov (2016).}:
\begin{equation}
    H(a)\equiv\dfrac{1}{a}\dfrac{da}{dt}=H_{0}\sqrt{\Omega_{\Lambda}+\Omega_{ \rm cdm}a^{-3}+\Omega_{\rm b}a^{-3}+\Omega_{\gamma}a^{-4}+\sum_{\nu}\Omega_{\nu}f_{\nu}(a)}
\end{equation}
Where $H_{0}$ is the present value of the Hubble parameter, $\Omega_{\rm cdm},\Omega_{\rm b},\Omega_{\gamma},\Omega_{\nu}$ and $\Omega_{\Lambda}$ are the current fractional energy densities of cold dark matter, baryons, photons and dark energy in the Universe, and the functions $f_{\nu}(a)$ determine the contributions of each of the four states of neutrinos (for which in expression (6) the there is a sum). The energy density $\rho_{\nu}$ of each is determined by the Fermi-Dirac distribution function:
\begin{equation}
    \rho_{\nu}(a,m,N_{\rm eff})=2\int_{0}^{\infty}\dfrac{4\pi p^2dp}{(2\pi\hbar)^3}\dfrac{\sqrt{(p{\cdot}c)^2+(m{\cdot}c^2)^2}}{exp\left(\dfrac{p{\cdot}c}{N_{\rm eff}^{1/4}kT(a)}\right)+1}~,~~ T(a)=\frac{T^{\,0}_{\scaleto{\mathrm{ C\nu B}}{4pt}}}{a}
\end{equation}
Here $m$ is the neutrino mass, $T^{\,0}_{\scaleto{\mathrm{ C\nu B}}{4pt}}=1.9454\,K$ is the present value of the relic neutrino temperature, defined by the following relation: $T^{\,0}_{\scaleto{\mathrm{ C\nu B}}{4pt}}=\left(\frac{4}{11}\right)^{\frac{1}{3}}T^{\,0}_{\rm \text{\tiny CMB}}$, $T(a)$ is the dependence of the neutrino temperature on the scale factor due to the expansion of the Universe, the multiplier $2$ before the integral takes into account the antineutrino contribution, $N_{\rm eff}$ is the effective number of relativistic degrees of freedom per neutrino state, which is usually introduced based on the relativistic neutrino energy density and not on the distribution function (7).  The parameterization used in this paper corresponds to the definition of $N_{\rm eff}$ of the numerical code ``CAMB'' (Lewis et al., 2000) used by the Planck Collaboration. The physical meaning of the parameter $N_{\rm eff}$ can be seen from the expression for the neutrino energy density $\rho_{\nu}^{\rm rel}$ in the relativistic limit:
\begin{equation}\rho_{\nu}^{\rm rel}=\dfrac{7}{8}\Bigl(\dfrac{4}{11}\Bigr)^{\frac{4}{3}}N_{\rm eff}\rho_{\gamma}=\dfrac{7}{8}\dfrac{\pi^2}{15}\dfrac{k^4}{(\hbar{\cdot}c)^3}N_{\rm eff}T^{4}(a)\end{equation}
The parameter $N_{\rm eff}$ indeed characterises the effective number of relativistic neutrinos, as it enters into expression (8) additively with respect to the number of their states. In the standard $\Lambda \rm CDM$ model the total value of $N_{\rm eff}$ is chosen to be three ($N_{\rm eff}^{\rm std}=3.046$) \footnote{The $0.046$ correction results from the additional heating of neutrinos during electron-positron annihilation (see, e.g., Mangano et al., 2005)} in accordance with the number of active neutrinos. In addition, $N_{\rm eff}$ can be used to parameterise the effects associated with the change $\Delta T^{\,0}_{\scaleto{\mathrm{ C\nu B}}{4pt}}$ in neutrino temperature value $T^{\,0}_{\scaleto{\mathrm{ C\nu B}}{4pt}}$: $N_{\rm eff}=3.046\bigl(1+\frac{\Delta T^{\,0}_{\scaleto{\mathrm{ C\nu B}}{4pt}}}{T^{\,0}_{\scaleto{\mathrm{ C\nu B}}{4pt}}}\bigr)^{4}$ . Therefore, equation (8) gives the relation between the neutrino temperature $T$, their effective number of relativistic degrees of freedom $N_{\rm eff}$ and the corresponding relativistic neutrino energy density $\rho_{\nu}^{\rm rel}$.

The expression $H(a)$ includes the neutrino energy density relative to the present value of the critical density $\rho_{\rm c}$
\begin{equation*}
   \dfrac{\rho_{\nu}(a,m,N_{\rm eff})}{\rho_{\rm c}}=\Omega_{\nu}(N_{\rm eff},m)f_{\nu}(a; N_{\rm eff},m)~~,~~\Omega_{\nu}=\dfrac{3\zeta(3)}{2\pi^2}\dfrac{(kT^{\,0}_{\scaleto{\mathrm{ C\nu B}}{4pt}})^3}{\hbar^3c\rho_{\rm c}}N_{\rm eff}^{\frac{3}{4}}m
   \end{equation*}
\begin{equation}
    f_{\nu}(a)=\dfrac{2}{3\zeta(3)}\dfrac{1}{a^4}\displaystyle\int_{0}^{\infty}\dfrac{y^2}{\exp(y)+1}\left(\sqrt{\left(\dfrac{kT^{\,0}_{\scaleto{\mathrm{ C\nu B}}{4pt}}}{c^2}\dfrac{N_{\rm eff}^{1/4}}{ma}\right)^2y^2+1}\right)dy
\end{equation}

where $\zeta(x)$ is the Riemann zeta function ($\zeta(3)\approx1.20206$), $\Omega_{\nu}$ is the relative neutrino energy density in the nonrelativistic limit. In the limit $\frac{mc^2}{kT}\gg1$, expression (9) contributes to the Hubble parameter as a relativistic degree of freedom, since $\Omega_{\nu}f_{\nu} \sim N_{\\rm eff}{\cdot}a^{-4}$, and in the opposing case $\frac{mc^2}{kT}\ll1$ -- as nonrelativistic: $\Omega_{\nu}f_{\nu} \sim N_{\rm eff}^{3/4}{\cdot m}{\cdot}a^{-3}$. Hence, the fourth neutrino can influence the expansion of the Universe during two key phases of its evolution: the radiation-dominated and matter-dominated eras.

The accounting of the sterile neutrino will also change the estimate $\rm H_{0}$ derived from the relic radiation anisotropy analysis using the angular acoustic scale $\theta_{*}$ at the time of recombination $t_{*}$:
\begin{equation*}
    \theta_{*}=\frac{R_{\rm s}}{D_{\rm a}}~~,
    \end{equation*}
    \begin{equation}
    R_{\rm s}=a(t_{*})\int_{0}^{t_{*}}\frac{c_{\rm s}(t)}{a(t)}dt ~,~D_{\rm a}=a(t_{*})\int_{t_{*}}^{t_{0}}\frac{cdt}{a(t)}
\end{equation}
Here $R_{\rm s}$ is the size of the sound horizon, determined using the speed of sound waves in the expanding early Universe $c_{\rm s}=\frac{c}{\sqrt{3}}/\sqrt{(1+\frac{\Omega_{\rm b}}{\Omega_{\gamma}}a)}$, $D_{\rm a}$ -- the distance travelled by CMB photons from the moment of recombination until the current time $t_{0}$ (angular diameter distance), $a(t_{*})\equiv a_{*}$ is the value of the scale factor at the moment of recombination (corresponding to the redshift $z_{*}=1089.92$, see Planck Collaboration, 2020). Explicitly, the values $R_{\rm s}$ and $D_{\rm a}$ are given by the following expressions:
\begin{equation}
    R_{\rm s}=\frac{c}{H_{0}\sqrt{3}}a(t_{*})\int_{0}^{a(t_{*})}\frac{da}{a^2\sqrt{(1+\frac{3\Omega_{\rm b}}{4\Omega_{\gamma}}a)(\Omega_{\Lambda}+(\Omega_{\rm b}+\Omega_{\rm  cdm})a^{-3}+\Omega_{\gamma}a^{-4}+\sum_{\nu}\Omega_{\nu}f_{\nu}(a))}}
\end{equation}
\begin{equation}
    D_{\rm a}=\frac{c}{H_{0}}a(t_{*})\int_{a(t_{*})}^{1}\frac{da}{a^2\sqrt{(\Omega_{\Lambda}+(\Omega_{\rm b}+\Omega_{\rm cdm})a^{-3}+\Omega_{\gamma}a^{-4}+\sum_{\nu}\Omega_{\nu}f_{\nu}(a))}}
\end{equation}
It may seem that the value $H_{0}$ is not included in the ratio $R_{\rm s}$ to $D_{\rm a}$ that determines $\theta_{*}$, but this is not the case because the definition of parameters $\Omega_{i}~,~i\in\{\Lambda,\rm cdm,\rm b,\nu\}$ includes critical density $\rho_{c}$, which itself depends on $H_{0}$. Introduction of the following standard notations
\begin{equation*}h=\frac{ H_{0}}{100 \text{km}{\cdot}\text{s}^{-1}{\cdot} \text{Mpc}^{-1}}~~,~~\omega_{i}=\Omega_{i}h^2~,~i\in\{\Lambda,\rm cdm , \rm b , \nu\}
\end{equation*}
allows the dependence $\theta_{*}$ on $ H_{0}$ (via $h$) to be written out explicitly:
\begin{equation}
    \theta_{*}=\frac{1}{\sqrt{3}}\ddfrac{\int_{0}^{a(t_{*)}}\dfrac{da}{\sqrt{(1+\frac{3\omega_{\rm b}}{4\omega_{\gamma}}a)((h^2-\omega_{\rm m})a^4+\omega_{\rm m}a+\omega_{\gamma}+\sum_{\nu}\omega_{\nu}f_{\nu}(a))}}}{\int_{a(t_{*})}^{1}\dfrac{da}{\sqrt{(h^2-\omega_{\rm m})a^4+\omega_{\rm m}a+\omega_{\gamma}+\sum_{\nu}\omega_{\nu}f_{\nu}(a)}}}
\end{equation}
Here $\omega_{\rm m}=\omega_{\rm b}+\omega_{\rm cdm}$. The parameter values $\theta_{*},\omega_{\rm b}, \omega_{\rm cdm}$ are derived from the relic radiation anisotropy analysis, the value of $\omega_{\gamma}$ is determined from the measured $\rm CMB$ Planck spectrum, characterised by the temperature $T^{\{\,0}_{\rm \text{\tiny CMB}}$, and the value of $\omega_{\nu}$ is obtained theoretically with the given parameters within the standard cosmological model. Therefore, the numerical solution to this equation allows the value of $\rm H_{0}$ to be determined.
\begin{figure}[t]
		\centering
		\includegraphics[width=0.85\textwidth]{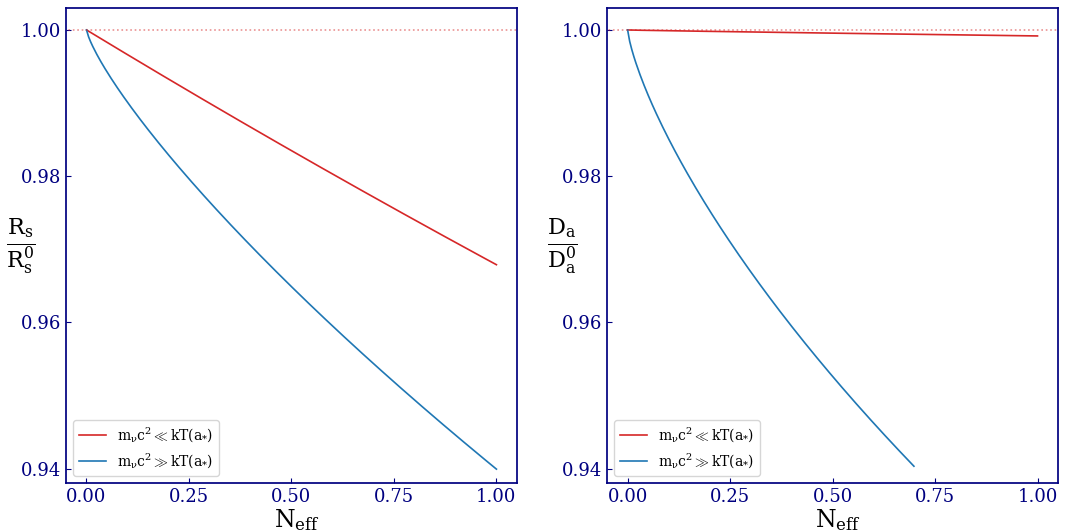}
		\caption{\textit{Left panel}: The dependence of the sound horizon $ R_{\rm s} $ on the effective number of neutrino species $ N_{\rm eff} $ in the case of neutrino with mass $ m_{\nu}c^2\ll kT(a_{*}) $ (red curve, assuming $m_{\nu}=~0. 01$\,eV as an example) and $ m_{\nu}c^2\gg kT(a_{*})$ (blue curve, assuming $m_{\nu}=~2.7$\,eV as an example) scaled to the standard value $ R_{\rm s}^{0}$ (Planck Collaboration, 2020). \textit{Right panel}: The dependence of the angular diameter distance $ D_{\rm a} $ on $ N_{\rm eff} $, scaled to the standard value $D_{\rm a}^{0}$.}
\end{figure}

\begin{figure}[t]
	 	\centering
	 	\includegraphics[width=0.85\textwidth]{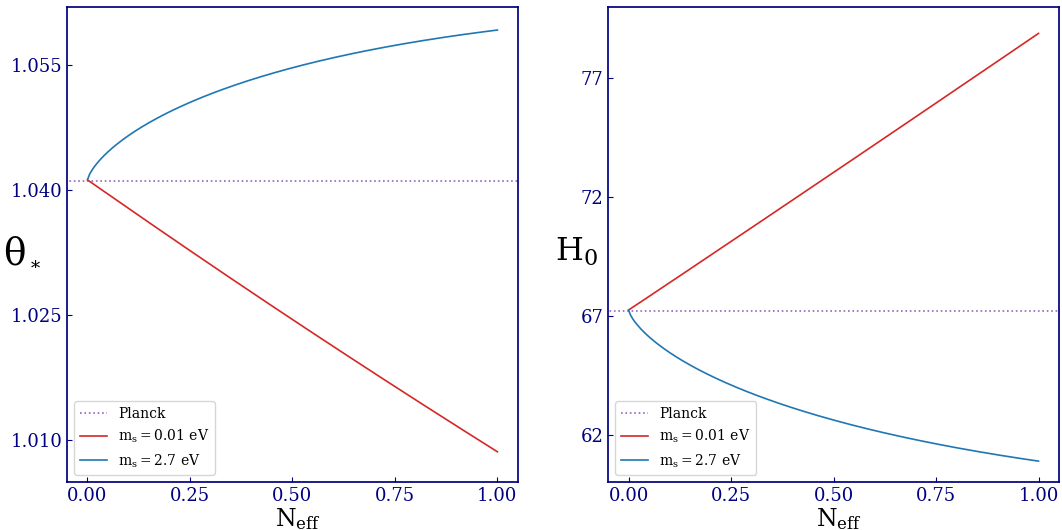}
	 	\caption{\textit{Left panel}: The dependence of angular acoustic scale $ \theta_{*} $ on $ N_{\rm eff} $ in the case of neutrino with mass $ m_{\nu}c^2\ll kT(a_{*}) $ (red curve, assuming $m_{\nu}=~0. 01$\,eV as an example) and $ m_{\nu}c^2\gg kT(a_{*})$ (blue curve, assuming $m_{\nu}=~2.7$\,eV as an example). The dotted line indicates the standard value (Planck Collaboration, 2020).
	 	\textit{Right panel}: The dependence of $ H_{0} $ on $ N_{\rm eff} $. The dotted line represents the standard value.}
	 \end{figure}
	 
The sound horizon $R_{\rm s}$, as can be seen from (11), is determined predominantly by the relativistic matter density, while the angular diameter distance $D_{\rm a}$ -- by the nonrelativistic matter density. The introduction of a fourth sort of neutrino will\footnote{As noted in the previous section, the addition of another neutrino to the standard $\Lambda \rm CDM$ model will affect the estimates of the main cosmological parameters (in particular, $\theta_{*}, \omega_{\rm cdm}, \omega_{\rm b}$). In this section their variations are not considered.}, other things being equal, lead to a decrease in both $R_{\rm s}$ and $D_{\rm a}$ (Figure 2), but the ratio of these quantities, that is $\theta_{*}$, can both increase and decrease, depending on which decreases faster -- $R_{\rm s}$ or $D_{\rm a}$. As can be seen from Figure 2, neutrinos with masses $m_{\nu}c^2\ll kT(a_{*})$ ($T(a_{*})=0.19\,$eV) have almost no effect on $D_{\rm a}$ compared to $R_{\rm s}$, leading to a decrease in their ratio $\theta_{*}$. In the case of $m_{\nu}c^2\gg kT(a_{*})$ the value $D_{\rm a}$, on the contrary, decreases much more than $R_{\rm s}$ and hence $\theta_{*}$ increases. The behaviour of $\theta_{*}$ in case these inequalities are met is shown in Figure 3. The value of $\theta_{*}$ is determined with high accuracy by the analysis of the CMB anisotropy, so the influence on its theoretical value (equation 13) of the fourth state of neutrino must be compensated by changes in the estimates of other cosmological parameters. Of all the parameters in formula (13), only the value of $H_{0}$ is not evaluated by the anisotropy analysis of $\rm CMB$, but is a derived parameter the change in which can compensate for the change in $\theta_{*}$. The value of $H_{0}$ is inversely proportional to both the sound horizon $R_{\rm s}$ and the angular diameter distance $D_{\rm a}$, but the latter is affected more, resulting in an inverse relation between $H_{0}$ and $\theta_{*}$ (Figure 3, right-hand panel). For clarity, all the variations described in this section are illustrated in Figure 4.

	 \begin{figure}[t]
	  \vskip -0.5cm
	 	\centering	 	\includegraphics[width=1\textwidth]{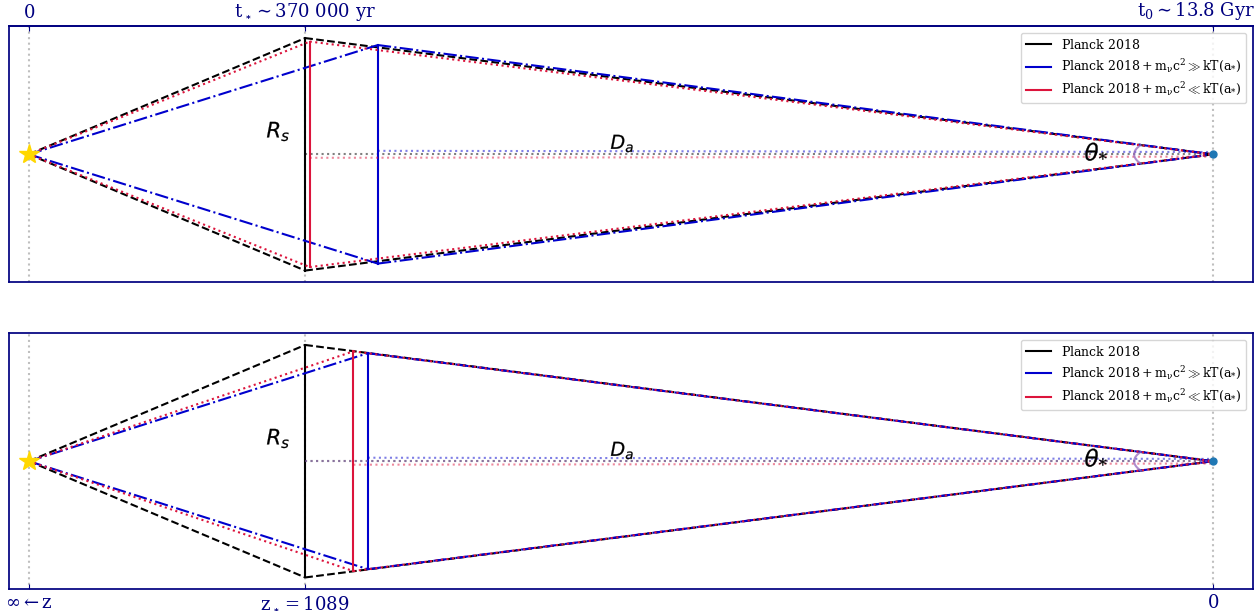}
	 	\caption{ Behaviour of the main scales determining the value of $\theta_{*}$ depending on the parameters of the fourth neutrino. Relative to the standard values of $R_{\rm s}$ and $D_{\rm a}$ (black vertical line and horizontal dashed line, respectively), their magnitudes are shown in the case of the fourth neutrino with $m_{\nu}c^2\gg kT(a_{*})$ (blue rhombus corresponding to $m_{\nu}=2. 7$ eV) and neutrino with $m_{\nu}c^2\ll kT(a_{*})$ ( red rhombus corresponding to $m_{\nu}=0.01$ eV).
	 	\textit{Top panel}: The values $R_{\rm s}$ and $D_{\rm a}$ are calculated assuming the standard values of cosmological parameters.
	 	\textit{Bottom panel}: The values $R_{\rm s}$ and $D_{\rm a}$ are calculated taking into account the CMB anisotropy analysis -- the main cosmological parameters are allowed to vary in each of the three cases.  $N_{\rm eff}$ for the fourth neutrino in both cases is taken to be 1.}
	 \end{figure}
	 
	 As a result, an increase in the effective number of neutrino species $N_{\rm eff}$ should lead to a decrease in the estimate of the present Hubble parameter value $H_{0}$ if the neutrino is massive enough ($m_{\nu}\gg0.19$ eV), and, conversely, to an increase in the estimate $H_{0}$ if it is light enough ($m_{\nu}\ll0.19$ eV). This statement will be checked numerically in the next section.

	 	\pagebreak
\section*{\large The method of determining cosmological parameters}
\vspace{-5mm}

The CMB anisotropy data (TT+TE+EE+lowE+lensing -- the measured values of the coefficients $D_{l}^{\text{\tiny TT}}$,$D_{l}^{\text{\tiny TE}}$,$D_{l}^{\text{\tiny EE}}$, polarisation degree of the $\rm CMB$ at low values of $l$ and gravitational lensing respectively) were taken from the Planck Collaboration open archive\footnote{https://pla.esac.esa.int
}.
 The code ``CAMB'' (Lewis et al., 2000) was used to solve the system of differential equations describing the evolution of CMB temperature fluctuations, and the code ``Cobaya'' (Torrado et al., 2021) was used to determine cosmological parameters. (These programs and their adaptations have been used by the Planck Collaboration.)

	When determining the cosmological parameters, neutrinos were accounted for in two cases: a sterile neutrino with mass $ m_{\rm s} = 1$ ev, corresponding to the results of several reactor experiments and a neutrino with mass $m_{\rm s} = 2.7$ eV, corresponding to the results of the ``Neutrino-4'' experiment, were considered. For each of these cases the calculation was carried out with $ N_{\rm eff}^s= 0.1,~0.5,~1$, and the corresponding value of the effective number of neutrino species for the active states was chosen to be standard -- $N_{\rm eff}^{a}=3.046$. The results of the calculations are given in Table 1.
	\newpage

	\begin{table}[h!]
		\centering
				\caption{The dependence of cosmological parameters on the effective number of sterile neutrinos with mass $ m_{\rm s}=1 $ eV (top table) and $m_{\rm s}=2.7$ eV (bottom table). The values of $\Omega_{\rm cdm},~\Omega_{\rm b},~\Omega_{\nu}$ are given as percentages and $H_{0}$ is given as units of $\small \frac{\text{km}}{\text{s}{\cdot\text{Mpc}}}$. The values of $\Omega_{\rm m}$ and $\Omega_{\Lambda}$ are defined as follows: $\Omega_{\rm m}=\Omega_{\rm cdm}+\Omega_{\rm b}+\Omega_{\nu}$, $\Omega_{\Lambda}=1-\Omega_{\rm m}$. The second column shows the standard values for all parameters.\newline}
				\vskip 0.5cm
				\begin{tabular}{|l|c| c c c|}
			\hline
			\multirow{2}{*}{$\rm Parameter$}& \multirow{2}{*}{$\rm{Planck 2018}$} &\multicolumn{3}{c|}{$\rm{m_{s} = 1 eV}$}  \\
			\cline{3-5}
			&  & $\rm{ N_{eff}^{s}=0.1}$ & $\rm{N_{eff}^{s}=0.5}$ & $\rm{ N_{eff}^{s}=1}$ \\
			\hline
			$~~~~~\Omega_{\rm cdm}$	& $\rm{26.45\pm0.50}$ & $\rm{27.07\pm0.51}$ & $\rm{29.06\pm0.56}$ & $\rm{30.84\pm0.64}$ \\
			\hline
			$~~~~~\Omega_{\rm b}$	& $\rm{~4.93\pm0.09}$ & $\rm{~5.01\pm0.09}$ & $\rm{~5.14\pm0.09}$ & $\rm{~5.17\pm0.35}$ \\
			\hline\hline
			$~~~~~\Omega_{\nu}$	&
			$\rm{0.14}$ & $\rm{0.57}$ & $\rm{1.59}$ & $\rm{2.57}$ \\
			\hline\hline
			$~~~~~\Omega_{\rm m}$	& $\rm{31.53\pm0.73}$ & $\rm{32.66\pm0.76}$ & $\rm{35.81\pm0.85}$ & $\rm{38.59\pm0.97}$ \\
			\hline
			$~~~~~\Omega_{\Lambda}$	& $\rm{68.47\pm0.73}$ & $\rm{67.34\pm0.76}$ & $\rm{68.42\pm0.85}$ & $\rm{61.41\pm0.97}$ \\
			\hline\hline
			$~~~~~\rm{H_{0}}$	& $\rm{67.36\pm0.54}$ & $\rm{66.91\pm0.53}$ & $\rm{66.24\pm0.54}$ & $\rm{66.28\pm0.57}$ \\
			\hline
		\end{tabular}
	\vskip 0.5cm
		\begin{tabular}{|l|c|c c c|}
			\hline
			\multirow{2}{*}{$\rm Parameter$}& \multirow{2}{*}{$\rm{Planck 2018}$}& \multicolumn{3}{c|}{$\rm{m_{s} = 2.7 eV}$} \\
			\cline{3-5}
			&  & $\rm{ N_{eff}^{s}=0.1}$ & $\rm{ N_{eff}^{s}=0.5}$ & $\rm{N_{eff}^{s}=1}$ \\
			\hline
			$~~~~~\Omega_{\rm cdm}$	& $\rm{26.45\pm0.50}$  & $\rm{26.79\pm0.51}$ & $\rm{29.35\pm0.56}$ & $\rm{32.36\pm 0.57}$ \\
			\hline
			$~~~~~\Omega_{\rm b}$	& $\rm{~4.93\pm0.09}$  & $\rm{~5.07\pm0.09}$ & $\rm{~5.49\pm0.10}$ & $\rm{~5.88\pm0.11}$ \\
			\hline\hline
			$~~~~~\Omega_{\nu}$	& $\rm{0.14}$  & $\rm{1.30}$ & $\rm{4.31}$ & $\rm{7.58}$ \\
			\hline \hline
			$~~~~~\Omega_{\rm m}$	& $\rm{31.53\pm0.73}$ & $\rm{33.18\pm0.77}$ & $\rm{39.18\pm0.91}$ & $\rm{45.8\pm1.1}$ \\
			\hline
			$~~~~~\Omega_{\Lambda}$	& $\rm{68.47\pm0.73}$ & $\rm{66.82\pm0.77}$ & $\rm{60.82\pm0.91}$ & $\rm{54.2\pm1.1}$ \\
			\hline\hline
			$~~~~~\rm{H_{0}}$	& $\rm{67.36\pm0.54}$  & $\rm{66.50\pm0.53}$ & $\rm{64.07\pm0.51}$ & $\rm{62.20\pm 0.53}$ \\
			\hline
		\end{tabular}
		\end{table}
		
	\newpage
 Neutrinos with masses $ m_{\rm s}=1 $ and $2.7~$eV turned out to be heavy enough for the present value of the Hubble parameter to start decreasing with the increasing value of $N_{\rm eff} $, as suggested in the previous section. Hence, the accounting of the sterile neutrino detected both in the ``Neutrino-4'' experiment ($m_{s}=2.7$ eV) and in the earlier reactor experiments ($m_{s}=1$ eV) exacerbates the ``$H_{0} $-tension'' problem. In both cases, one also observes a change in other cosmological parameters determining the expansion rate of the Universe: in particular, the relationship between $\Omega_{m}$ and $\Omega_{\Lambda}$, which determines the behaviour of $H(t)$ in the later stages of its evolution, has changed significantly. Changes in the $H(t)$ relation will be discussed in more detail in the next section.
 
\subsection*{{\large Change in the value of $T_{\scaleto{\rm C\nu B}{4pt}}$.}}
\vspace{-5mm}

In addition to investigating the effects arising from the accounting for sterile neutrino, we studied the effects originating from the increase in the present value of the temperature of active states $T^{\,0}_{\scaleto{\rm C\nu B}{4pt}}=\Bigl(\dfrac{4}{11}\Bigr)^{\frac{1}{3}}T_{\gamma}^{0}=1.95\,K$ on cosmological parameter estimates, which is equivalent to increasing their effective number of species $N_{\rm eff}$.

There is a broad range of dark matter decay and annihilation processes that produce active neutrinos (Cirelli et al., 2011). Since the mass of cold dark matter particles should be much larger than the mass of active states ($m_{\nu}\lesssim1\,$eV), the neutrinos created during such reactions will be relativistic, and their corresponding energy density correction $\Delta\rho_{\nu}$ can be expressed in terms of either $N_{\rm eff}$ or $T^{\,0}_{\scaleto{\rm C\nu B}{4pt}}$ using expression (8). Therefore, dark matter decay and annihilation processes can lead to an increase in the parameter $N_{\rm eff}$ (or, equivalently, an increase in $T^{\,0}_{\scaleto{\rm C\nu B}{4pt}}$) of neutrino active states. A similar effect can be caused by the decay of a second, heavier sterile state with $m_{s}\sim1\,$keV. The existence of heavy sterile neutrinos is required to explain the process of active neutrino mass generation via the seesaw mechanism (Minkowski, 1997). Additionally, such neutrinos can also be introduced to explain the nature of dark matter, in which case the current observational data on the primordial nucleosynthesis, the background relic and gamma-ray radiation limit the allowed parameter region of these neutrinos ($m_{s}\sim1\,$keV, $\theta_{s}^2<10^{-5}$, see, for example, Boyarsky et al., 2009). Alternatively, the mentioned sterile neutrinos could exist in the universe independently of their contribution to dark matter (i.e. their number and energy density is only a small fraction of dark matter), which would significantly relax the existing parameter restrictions, while their decay would still lead to an increase in active neutrino density. This happens due to the following processes. The mixing of sterile neutrinos with active ones leads to decay of mass states via two reactions: into three lighter neutrinos ($\nu_{4}\to3\nu_{a}$), or into one lighter state with emission of a photon ($\nu_{4}\to\gamma+\nu_{a}$). The lifetime of neutrinos in both processes is substantially longer than the age of the Universe (see, e.g., Gorbunov, 2014, Dasgupta et al., 2021), while the decay of even a small number of such massive neutrinos can significantly alter $N_{\rm eff}$ of the active states. For example, a single sterile neutrino of mass $1\,keV$ would decay into an active neutrino with energy of $\sim 1/3\,\text{keV}$, which is equivalent to the energy density of $\sim10^4$ relic neutrinos (with a temperature of $1.95\,$K).

In the presented work the influence of the sterile neutrino decay processes on cosmological parameters is taken into account by increasing $T_{\scaleto{\rm C\nu B}{4pt}}$ (equivalent to increasing $N_{\rm eff}$) of the active states. A more precise calculation will be done in the subsequent works. First, the case of existence of only three active states was investigated, the temperature $T_{\scaleto{\rm C\nu B}{4pt}}$ of each varying in the range from $1.95\,\rm K$ to $2.07\,\rm K$ with step $0.03\,\rm K$. The results of the corresponding calculations are shown in Table 2.
 
A rise in $T^{\,0}_{\scaleto{\mathrm{ C\nu B}}{4pt}}$ results in a significant increase in the present value of the Hubble parameter, up to an agreement with the observational estimate $H_{0}=73.04\pm1.04~\text{km}\cdot\text{s}^{-1}\cdot \text{Mpc}^{-1}$ (Riess et al, 2022) at $T^{\,0}_{\scaleto{\mathrm{ C\nu B}}{4pt}}=2.07~\rm K$ (in this case $H_{0}=72.81^{+0.62}_{-0.56}~\text{km}\cdot\text{s}^{-1}\cdot \text{Mpc}^{-1}$). This is caused by an increase in the values of $N_{\rm eff}^{a}$ ($N_{\rm eff}^{a}=\bigl(1+\frac{\Delta T^{0}_{\scaleto{\mathrm{ C\nu B}}{4pt}}}{T^{\,0}_{\scaleto{\mathrm{ C\nu B}}{4pt}}}\bigr)^{4}$), corresponding to the three active states, since their mass is substantially less than $0.19\,$eV. The values of the other parameters influencing the expansion rate of the Universe do not differ as much from the standard ones, as in the case of the sterile neutrino. 

The results of accounting for the two effects simultaneously -- temperature increase $T^{\,0}_{\scaleto{\mathrm{ C\nu B}}{4pt}}$ for the model with sterile neutrino, are given in Table 3. The increase of the current Hubble parameter value $H_{0}$, caused by increasing $T^{\,0}_{\scaleto{\mathrm{ C\nu B}}{4pt}}$, is in this case compensated by its decrease due to an introduction of a sterile neutrino into the model.

\newpage
	\begin{table}[t]
		\centering
			\caption{The dependence of cosmological parameters on the present value of neutrino temperature $T^{\,0}_{\scaleto{\mathrm{ C\nu B}}{4pt}}$. The top row shows $T^{\,0}_{\scaleto{\mathrm{ C\nu B}}{4pt}}$ and its corresponding total number of effective neutrino species $N_{\rm eff}^{a}$. The values of $\Omega_{\rm cdm},~\Omega_{\rm b},~\Omega_{\nu}$ are given as percentages and $H_{0}$ is given as units of $\small \frac{\text{km}}{\text{s}{\cdot\text{Mpc}}}$. The values of $\Omega_{\rm m}$ and $\Omega_{\Lambda}$ are defined as follows: $\Omega_{\rm m}=\Omega_{\rm cdm}+\Omega_{\rm b}+\Omega_{\nu}$, $\Omega_{\Lambda}=1-\Omega_{\rm m}$. The second column shows the standard values for all parameters.\newline}
		\resizebox{\textwidth}{!}{%
		\begin{tabular}{|l|c|c c c c|}
			\hline
			\multirow{2}{*}{$\rm Parameter$}& \multirow{2}{*}{$\rm{Planck 2018}$}& $\rm{T^{\,0}_{\scaleto{\mathrm{ C\nu B}}{4pt}}=1.98~K~,}$&$\rm{T^{\,0}_{\scaleto{\mathrm{ C\nu B}}{4pt}}=2.01~K~,}$&$\rm{T^{\,0}_{\scaleto{\mathrm{ C\nu B}}{4pt}}=2.04~K~,}$&$\rm{T^{\,0}_{\scaleto{\mathrm{ C\nu B}}{4pt}}=2.07~K~,}$ \\
			
			&  &$N_{\rm eff}^{a}=3.3$ & $N_{\rm eff}^{a}=3.5$ & $N_{\rm eff}^{a}=3.7$ & $N_{\rm eff}^{a}=3.9$\\
			\hline
			$~~~~~\Omega_{\rm cdm}$	& $\rm{26.45\pm0.50}$ & $\rm{26.01\pm0.49}$ & $\rm{25.65\pm0.49}$ & $\rm{25.28\pm0.52}$ &
			$\rm{24.92\pm0.49}$\\
			\hline
			$~~~~~\Omega_{\rm b}$	& $\rm{~4.93\pm0.09}$ & $\rm{~4.76\pm0.08}$ & $\rm{~4.62\pm0.08}$ & $\rm{~4.47\pm0.08}$ &
			$\rm{~4.33\pm0.08}$\\
			\hline\hline
			$~~~~~\Omega_{\nu}$	& $\rm{0.14}$ & $\rm{0.14}$ & $\rm{0.14}$ & $\rm{0.15}$ &
			$\rm{0.15}$\\
			\hline\hline
			$~~~~~\Omega_{\rm m}$	& $\rm{31.53\pm0.73}$ & $\rm{30.93\pm0.73}$ & $\rm{30.42\pm0.72}$ & $\rm{29.91\pm0.75}$ &
			$\rm{29.41\pm0.87}$\\
			\hline
			$~~~~~\Omega_{\Lambda}$	& $\rm{68.47\pm0.73}$  & $\rm{69.06\pm0.73}$ & $\rm{69.57\pm0.72}$ & $\rm{70.09\pm0.75}$ &
			$\rm{70.58\pm0.87}$\\
			\hline\hline
			$~~~~~\rm{H_{0}}$	& $\rm{67.36\pm0.54}$ & $\rm{68.80\pm0.56}$ & $\rm{70.09\pm0.57}$ & $\rm{71.43\pm0.62}$ &
			$\rm{72.81^{+0.62}_{-0.56}}$\\
			\hline
		\end{tabular}%
	}
\end{table}
	\begin{table}[h!]
		\centering
			\caption{The dependence of cosmological parameters on the present neutrino temperature $T^{\,0}_{\scaleto{\mathrm{ C\nu B}}{4pt}}$ with a sterile state of mass $2.7$ eV. The first row shows the value of $T^{\,0}_{\scaleto{\mathrm{ C\nu B}}{4pt}}$ as well as the effective number of sterile neutrinos $N_{\rm eff}^{\rm s}$. The total number of effective degrees of freedom is given in the last row of the table.\newline}
			\resizebox{0.46\textwidth}{!}{%
			\begin{tabular}{|l|c|c|}
			\hline
			\multirow{2}{*}{$\rm Parameter$}& \multirow{2}{*}{$\rm{Planck 2018}$}& $\rm{T^{\,0}_{\scaleto{\mathrm{ C\nu B}}{4pt}}=2.07~K~,}$ \\
			
			&  &$ N_{\rm eff}^s=1$\\
			\hline
			$~~~~~\Omega_{\rm cdm}$	& $\rm{26.45\pm0.50}$ & $\rm{31.17\pm0.73}$ \\
			\hline
			$~~~~~\Omega_{\rm b}$	& $\rm{~4.93\pm0.09}$ & $\rm{~5.22\pm0.10}$ \\
			\hline\hline
			$~~~~~\Omega_{\nu}$	& 
			$\rm{0.14}$ & 
			$\rm{7.88}$  \\
			\hline\hline
			$~~~~~\Omega_{\rm m}$	& $\rm{31.53\pm0.73}$  & $\rm{44.3\pm1.2}$ \\
			\hline
			$~~~~~\Omega_{\Lambda}$	& $\rm{68.47\pm0.73}$  & $\rm{55.7\pm1.2}$ \\
			\hline\hline
			$~~~~~\rm{H_{0}}$	& $\rm{67.36\pm0.54}$ & $\rm{67.00\pm0.62}$ \\
			\hline
			\hline
			$~~~~~N_{\rm eff}$ & $3.046$ & $5.186$ \\
			\hline
		\end{tabular}%
		}
\end{table}

\newpage
\section*{\large The changes in the dependency $\boldsymbol{H(z)}$}
\vspace{-5mm}

The estimates of cosmological parameters derived from the CMB anisotropy analysis (their values are presented in Table 1) are used to plot the dependence of the Hubble parameter on the redshift $z$:
	\begin{equation}
	 H(z)=H_{0}\sqrt{\Omega_{\Lambda}+(\Omega_{\rm cdm}+\Omega_{\rm b})(1+z)^3+\Omega_{\gamma}(1+z)^4+\Omega_{\nu}(z)}
	\end{equation}
	 Here the neutrino contribution $\Omega_{\nu}=\Omega_{\rm a}(z)+\Omega_{\rm a}^{\rm m}(z)+\Omega_{s}(z)$ consists of three terms corresponding to two massless active states, one massive state with $m_{a}=0.06\,$eV and a sterile state with $m_{s}=2.7\,$eV respectively. Their explicit form is given by expression (9). Figure 5 shows the relative deviation of the Hubble parameter
	 \begin{equation}
	 \delta H(z)=\dfrac{H(z)-H_{\rm std}(z)}{H_{\rm std}(z)}\end{equation}
	 Here $H_{\rm std}(z)$ is calculated from standard values of cosmological parameters. The same figure shows the evolution of each term under the root in expression (14). In addition, Figure 6 shows the dependence of the total effective number of relativistic neutrinos $ N_{\rm eff} $.
	 
	 \newpage
	\begin{figure}[h!]
		\centering
		\includegraphics[width=0.85\textwidth]{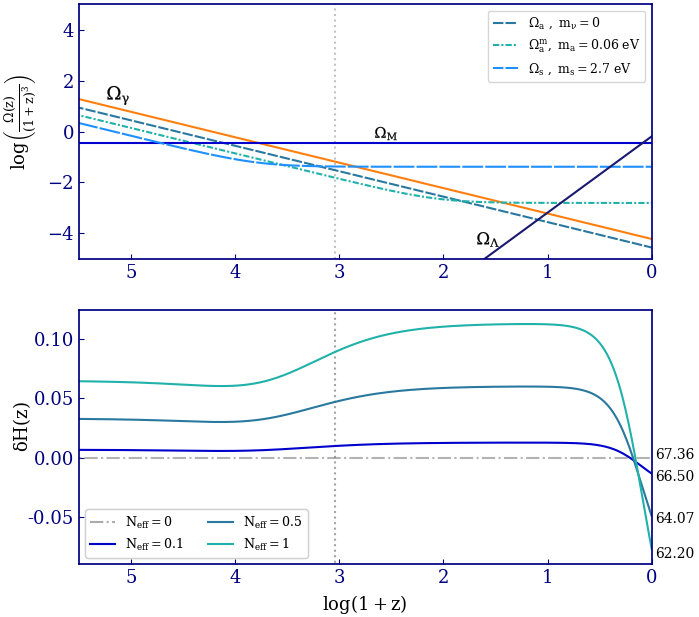}
	\caption{\newline \textit{Top panel}: The dependence of the relative energy densities of the components of the Universe corresponding to the terms under the root of expression (14) on the redshift for the case $ m_{\rm s}=2.7~ $eV. All energy densities are additionally divided by $(1+z)^3$. The vertical dashed line corresponds to the time of recombination ($z_{*}=1090$, Planck Collaboration, 2020).
		\textit{Bottom panel}:
		The relative deviation of the Hubble parameter from the standard behaviour as a function of the redshift for the case $ m_{\rm s}=2.7 ~$eV. The function $ H(z) $ at $ z\rightarrow0 $ gradually approaches its corresponding value $ H_{0} $ (see Table 1) for all values of $ N_{\rm eff} $ considered.}
	\end{figure}
	\begin{figure}[t]
	\centering
	\includegraphics[width=0.75\textwidth]{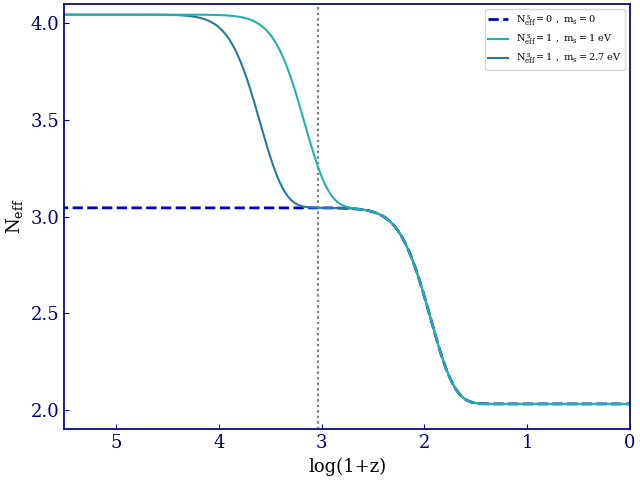}
	\caption{
	 The total number of relativistic neutrinos as a function of redshift for $m_{s}=1$ and $2.7\,$eV. $N_{\rm eff}$ is taken to be $1$ in both cases. The blue dashed line shows the standard dependence (in the model without the sterile neutrino). The vertical dashed line indicates the time of recombination. }
\end{figure}
\newpage

From the graphs in the top panel of Figure 5 for a given value of $ z $ it is possible to identify the form of matter that gives the dominant contribution to the Hubble parameter. Also, the moment of neutrino transition from relativistic to non-relativistic state is visible. As can be seen, for the heavy sterile state ($m_{\rm s}=2.7\,\text{eV}>0.19\,$eV) this transition occurs before recombination, and for the light active neutrino ($m_{\rm s}=0.06\,\text{eV}<0.19\,$eV) it happens after.

At values $z\gtrsim10^{3.5}$ the dominant contribution to $H(z)$ is relativistic matter, the density of which in the model with the sterile neutrino is much larger than the standard value. For this reason, the value of $H(z)$ is also larger, as can be seen in the bottom panel of figure 5. Around the value $z\sim10^{3. 5}$ the dominant contribution to $H(z)$ switches to nonrelativistic matter, including not only $\Omega_{\rm b}$ and $\Omega_{\rm cdm}$ (as in the standard case), but also the contribution of sterile neutrino in the nonrelativistic state $\Omega_{s}$ and a little later, around $z\sim10^2$, also the contribution of massive active neutrino $\Omega_{a}^{m}$ (the moments of massive neutrino transitions from the relativistic to nonrelativistic states can be clearly seen in Fig. 6). This leads to an even larger relative deviation $\delta H(z)$ for the redshift interval $z\sim10^{3.5}\div0.6$. At values $z\sim0.6$ a transition occurs from a decelerated expansion of the Universe to an accelerated one. The corresponding value of $z$ is determined from the equality $\Omega_{M}(1+z)^3=2\Omega_{\Lambda}$, therefore an increase in $\Omega_{M}$ would mean that the Universe had been expanding with deceleration longer, resulting in a lower value of $H_{0}$ compared to the standard scenario. This explains the decline in the $\delta H(z)$ dependence at $z\sim0.6\div0$.

The dependence $\delta H(z)$ behaves similarly when sterile neutrinos with $m_{s} = 1\,$eV are considered, however, since such neutrinos are lighter, their transition out of the relativistic state ends later, around values $ z\sim10^3 $, and partially overlaps with recombination (see Figure 6). 

Thus, consideration of the sterile neutrino leads to a significant change in the Hubble parameter during the entire evolution of the Universe, not just to a change in its present value.
 
\section*{\large Conclusion}
\vspace{-5mm}

In this paper an analysis of the CMB temperature anisotropy has been carried out, taking into account the possible existence of a sterile neutrino with masses $ 1 $ and $ 2.7$ eV. Estimates for the main cosmological parameters have been obtained. Including the Hubble parameter, the value of which in the two cases considered $ m_{\rm s}=1 ~$eV and $ m_{\rm s}=2.7~ $eV decreases with the increase of $ N_{\rm eff} $. The reason for such behaviour has been stated and analysed: introduction of a sterile neutrino into the standard $\Lambda \rm CDM $ model leads to a decrease of the angular diameter distance $ D_{\rm a} $ in the dependence $\theta_{*}(H_{0}) $, which turns out to be a determining factor. Based on the results of CMB anisotropy analysis, the $ H(z)$ dependence was reconstructed and its differences from the standard dependence were investigated in detail.

Apart from the effects associated with sterile neutrinos, we also considered the impact of a change in the effective number of neutrino species $N_{\rm eff}$ for the active states on the determination of cosmological parameters, which is equivalent to a change in their present temperature value $T^{\,0}_{\scaleto{\mathrm{ C\nu B}}{4pt}}$. An increase of $T^{\,0}_{\scaleto{\mathrm{ C\nu B}}{4pt}}$ has resulted in an increase of the present value of the Hubble parameter leading to its intersection with the observational estimate (2) at $T^{\,0}_{\scaleto{\mathrm{ C\nu B}}{4pt}}=2.07$ K, while not radically changing other cosmological parameters.

A decrease in the value of the Hubble parameter means an exacerbation of the ``$H_{0} $-tension'' problem. The results of this paper show that the introduction of a sterile neutrino into the standard $\Lambda \rm CDM $ model leads to an even larger discrepancy between the model-dependent estimate of the parameter $ H_{0} $ and its observed value. 

Thus, the potential existence of a sterile neutrino of mass $ 1 $ or $2.7$ eV would require a revision of the standard $\Lambda \rm CDM$ model to account for some new physics. On the other hand, effects leading to neutrino heating, such as the decay of sterile states or dark matter, could weaken or even eliminate the discrepancy between the estimates of $H_{0}$.

\subsection*{\large Acknowledgements}
\vspace{-5mm}

	This work was supported by a grant from the Russian Science Foundation №18-12-00301.
 
\pagebreak 

\subsection*{\large References}
\vspace{-5mm}
{\small
\noindent N. Allemandou, H. Almazan, P. del Amo Sanchez \etal, Journal of Instrumentation {\bf 13}, 07 (2018).\\
J. Ashenfelter, A.B. Balantekin, C. Baldenegro \etal, Nucl. Instrum. Meth. {\bf A922}, 287 (2019).\\
S.A. Balashev, E.O. Zavarygin, A.V. Ivanchik \etal, \mnras {\bf 458}, 2188 (2016).\\
V.V. Barinov, B.T. Cleveland, S.N. Danshin \etal, Phys. Rev. C {\bf105}, 065502 (2022).\\
S.M. Bilenky, B. Pontecorvo, Sov. J. Nucl. Phys. {\bf 24}, 316 (1976).\\
A. Boyarsky, O. Ruchayskiy, M. Shaposhnikov, Ann. Rev. Nucl. Part. Sci. {\bf 59}, 191 (2009).\\
M. Cirelli, G. Corcella, A. Hektor \etal, JCAP 1103:051 (2011).\\
M. Danilov, eprint arXiv:2012.10255 (2020).\\
S. Dasgupta, J. Kopp, Phys. Rept. {\bf 928}, 1 (2021).\\
D.J. Fixen, \apj\ {\bf 707}, 916 (2009).\\
S. Gariazzo, C. Giunti, M. Laveder \etal, J. High Energ. Phys. {\bf 6}, 135 (2017).\\
D.S Gorbunov, UFN {\bf 184}, 545 (2014).\\
Gorbunov D.S., Rubakov V.A, {\it Introduction to the Theory of the Early Universe: Hot Big Bang Theory} (M.: LENAND, 2016), vol.1\\
Gorbunov D.S., Rubakov V.A, {\it Introduction to the Theory of the Early Universe: Cosmological Perturbations and Inflationary Theory} (M.: LENAND, 2016), vol.2\\
A.V. Ivanchik, V.Yu. Yurchenko, Phys. Rev. D{\bf 98}, 081301 (2018).\\
O.A. Kurichin, P.A. Kislitsyn, V.V. Klimenko, et al., \mnras {\bf 502}, 3045 (2021).\\
Kurichin O.A, Kislitsyn P.A, Ivanchik A.V., \pazh {\bf 47}, 697 (2021).\\
O. Lahav, A.R Liddle, eprint arXiv:1912.03687 (2019).\\
A. Lewis, A. Challinor, A. Lasenby, Astrophys.J. {\bf 538}, 473 (2000).\\
G. Mangano, G. Miele, S. pastor \etal, Nucl.Phys. B{\bf 729}, 221 (2005).\\
G. Mention, M. Fechner, Th. Lasserre \etal, \prd {\bf 83}, 073006 (2011).\\
P. Minkowski, Phys. Lett. B {\bf 67}, 421 (1997).\\
Th. A. Mueller, D. Lhuillier, M. Fallot \etal, \prc {\bf 83}, 054615 (2011).\\
NEOS Collaboration, \prl {\bf 118}, 121802 (2017).\\
Planck Collaboration, \aap\ {\bf A6}, 641 (2020).\\
A.G. Riess, W. Yuan, L.M. Macri \etal, eprint arXiv:2112.04510 (2022).\\
A.P. Serebrov, R.M. Samoilov, V.G. Ivochkin \etal, \prd {\bf 104}, 032003 (2021).\\
SNO Collaboration, \prc\ {\bf 88}, 025501 (2013).\\
Super-Kamiokande Collaboration, \prl\ {\bf 82}, 2430 (1999).\\
J. Torrado and A. Lewis, JCAP {\bf 05}, 057 (2021).\\
Weinberg S., {\it Cosmology} (New York: Oxford University Press, 2008).\\
V.Yu. Yurchenko, A.V. Ivanchik, Astroparticle Physics {\bf 127}, 102537 (2021).\\
Zyla, (Particle Data Group), Prog. Theor. Exp. Phys.2020, {\bf 083C01} (2020).\\
}

\end{document}